\documentclass[aps,twocolumn,superscriptaddress]{revtex4-2}
\usepackage{graphicx}
\usepackage{bm}
\usepackage{bbold}
\usepackage{amsmath}

\usepackage{epsfig}
\usepackage{physics}
\usepackage{epstopdf}
\usepackage[driverfallback = dvips]{hyperref}
\usepackage{url}
\usepackage{color}
\usepackage{multirow}
\usepackage{comment}
\usepackage{mathrsfs}
\begin{document}
\title{Crystal structures and high-temperature superconductivity in molybdenum-hydrogen binary system under high pressure}

\author{Aiqin Yang}
\affiliation{MOE Key Laboratory for Non-equilibrium Synthesis and Modulation of Condensed Matter, Shaanxi Province Key Laboratory of Advanced Functional Materials and Mesoscopic Physics, School of Physics, Xi'an Jiaotong University, 710049, Xi'an, Shaanxi, P.R.China}
\author{Xiangru Tao}
\affiliation{MOE Key Laboratory for Non-equilibrium Synthesis and Modulation of Condensed Matter, Shaanxi Province Key Laboratory of Advanced Functional Materials and Mesoscopic Physics, School of Physics, Xi'an Jiaotong University, 710049, Xi'an, Shaanxi, P.R.China}
\author{Yundi Quan}
\affiliation{MOE Key Laboratory for Non-equilibrium Synthesis and Modulation of Condensed Matter, Shaanxi Province Key Laboratory of Advanced Functional Materials and Mesoscopic Physics, School of Physics, Xi'an Jiaotong University, 710049, Xi'an, Shaanxi, P.R.China}
\author{Peng Zhang}
\email{zpantz@mail.xjtu.edu.cn}
\affiliation{MOE Key Laboratory for Non-equilibrium Synthesis and Modulation of Condensed Matter, Shaanxi Province Key Laboratory of Advanced Functional Materials and Mesoscopic Physics, School of Physics, Xi'an Jiaotong University, 710049, Xi'an, Shaanxi, P.R.China}

\begin{abstract}
Motivated by advances in hydrogen-rich superconductors in the past decades, we conducted variable-composition structural searches in Mo-H binary system at high pressure. A new composition-pressure phase diagram of thermodynamically stable structures has been derived. Besides all previously discovered superconducting molybdenum hydrides, we also identified series of thermodynamically metastable superconducting structures, including I4/mmm-Mo$_3$H$_{14}$, I4cm-MoH$_9$, P4/nmm-MoH$_{10}$ and P42$_1$2-MoH$_{10}$, with the superconducting transition temperatures from 55 to 126 K at 300 GPa. In these superconducting molybdenum hydrides, vibrations of the Mo-atoms contributes significantly to the electron-phonon coupling and the superconducting transition temperature, in complementary to the contributions by the vibrations of the H-atoms. Our works highlight the importance of compounds with non-integer composition ratio and metastable states in material searches, for example the potential high temperature superconductors.
\end{abstract}
\maketitle

\section{Introduction}
In 1911, Onnes discovered superconductivity in mercury, which started an new era in science. \cite{Onnes1911} However, the progress in searching for high-temperature superconductors is limited. Even after more than a century of exploration, the highest superconducting transition temperature at ambient pressure is 133 K in cuprates \cite{Cu133K.nature1993}, and that of conventional superconductors is only 39 K in MgB$_2$ \cite{MgB2.nature2001}. 
Early in 1968, Ashcroft suggested that metallic hydrogen could be a candidate for high temperature superconductors due to its high-frequency phonons and strong electron-phonon coupling. \cite{Ashcroft.PRL1968} Unfortunately, the synthesizing of metallic hydrogen by physical compression requires extremely high pressure, which leads to ambiguities and discrepancies in experiments. \cite{Hydrogen.nphys2007,Hydrogen.PRB2012,Hydrogen.PRL2015,Hydrogen.nature2016,Hydrogen.PRL2018} In 2004, Ashcroft further proposed an alternative that pre-compression of hydrogen atoms by chemical pressure from surrounding atoms in hydrides could reduce the required physical pressure for superconductivity. \cite{Ashcroft.PRL2004} Following the strategy, series of high-temperature superconducting hydrides have been found in the last decade. The superconducting temperature of H$_3$S is 203 K at 155 GPa \cite{Duan.SciRept2014,Drozdov.Nature2015}, and that of LaH$_{10}$ is 260 K at 180-200 GPa. \cite{Peng.PRL2017,Liu.PNAS2017,Geballe.Angew2018,Somayazulu.PRL2019, Drozdov.Nature2019,Semenok.AM2022} Very recently, Dasenbrock-Gammon {\it et al.} \cite{Dias.Nature2023} even claimed that lutetium-nitrogen-hydrogen has $T_{\text{c}}$ of 294 K at 1 GPa, although most of the subsequent works cannot reproduce their results. \cite{Shan.CPL2023, Ming.arxiv2023, Xing.arxiv2023, Cai.arxiv2023, Salke.arxiv2023, DPeng.arxiv2023, Tao.SciBull2023} The efforts in searching for new superconducting hydrides are still on-going.

Molybdenum, crystallized in a body centered cubic structure under ambient conditions, is a typical transition metal with high melting point and toughness. \cite{Molybdenum2003} As early as 2003, the pressure-temperature phase diagram of the Mo-H binary system was reported by Fukai et al. \cite{Fukai.MT2003}. Several follow-up works have studied the crystal structure phase transitions in this system as well. \cite{Antonov.JPCM2004,Abramov.JAC2016,Kuzovnikov.JAC2017} 
The superconducting transition temperatures in MoH and MoD of P6$_3$/mmc structure are 0.92 K and 1.11 K at ambient pressure. \cite{Antonov1988} The hydrogen content in molybdenum hydrides will increase under compression. \cite{Fukai.MT2003,Kuzovnikov.JAC2017} Fukai et al. \cite{Fukai.MT2003} showed that the H/Mo ratio in the body-centered cubic $\alpha$-phase of Mo-H system is small at the ambient pressure, which increases significantly at high pressures to a stoichiometeric composition of around 1 in the hexagonal close packed $\epsilon$-phase or face-centered cubic $\gamma$-phase.

The density functional theory (DFT) based crystal structure prediction algorithms \cite{AIRSS,CALYPSO,USPEX} are proved successful in searching for superconducting hydrides, including the Mo-H binary superconductors. In 2016, Feng et al. \cite{Feng.SSC2016} investigated the MoH$_n$ (n = 1, 2 and 3) compounds below 100 GPa using the crystal structure prediction method. MoH$_2$ (P6$_3$mc) of the hexagonal structure, with the same arrangement of metal atoms as MoH (P6$_3$/mmc), was predicted stable at the pressure of 9-24 GPa, then transformed into orthorhombic structure (Pnma) under higher pressure. In 2020, Liao et al. \cite{Liao.JAP2020} conducted structural searches of MoH$_n$ (n = 1-6) up to 300 GPa, and found four new superconducting phases including I4/m-MoH$_4$ (T$_c$ of 41-49 K at 250 GPa), I4/mmm-MoH$_4$ (T$_c$ of 64-73 K at 300 GPa), C2/m-MoH$_6$ (T$_c$ of 3.1-5.9 K at 150 GPa), and Immm-MoH$_6$ (T$_c$ of 74-82 K at 300 GPa). In 2021, Du et al. \cite{Duan.PCCP2021} investigated the phases of MoH$_{n}$ (n=1-12) compounds at pressures up to 300 GPa, where three new superconducting phase Cmcm-MoH$_5$ (T$_c$ of 67-72 K at 150 GPa), Pnma-MoH$_6$ (T$_c$ of 22-27 K at 100 GPa) and Cmmm-MoH$_{11}$ (T$_c$ of 107-117 K at 250 GPa) were predicted.

The structural searches above are limited to compounds of MoH$_n$-type stoichiometry, while the potentially stable or superconducting Mo-H compounds of Mo$_m$H$_n$-type stoichiometry are missing. Actually, binary compounds of m:n composition ratio could be very important. Recent work by Tikhonov et al. \cite{TIKHONOV2023} reported thermodynamically stable vanadium hydrides of non-integer composition ratio, including I4/mmm-V$_3$H$_2$, C222-V$_6$H$_5$, R$\bar{3}$m-V$_3$H$_7$, Ibam-V$_2$H$_5$ and I$\bar{4}$2m-V$_4$H$_{11}$, using the variable-composition crystal structure prediction method. Another example is the lutetium hydrides. Li et al. \cite{Lu4H23_2023} discovered superconductivity of $T_{\text{c}}$ $\approx$ 71 K at 218 GPa in lutetium polyhydrides, with candidate of Pm$\bar{3}$n-Lu$_4$H$_{23}$ stoichiometry. 

In this work, we carried out comprehensive variable-composition structural searches of the Mo-H binary system up to 300 GPa. An updated pressure-composition phase diagram of Mo-H systems is derived, including five unprecedented molybdenum hydrides of non-integer H/Mo ratios. Besides, we also discovered series of thermodynamically metastable high-temperature superconducting structures, including I4/mmm-Mo$_3$H$_{14}$, I4cm-MoH$_9$, P4/nmm-MoH$_{10}$ and P42$_1$2-MoH$_{10}$, with estimated T$_c$ from 55 to 126 K at high pressure. Therefore, our works established an example to stress the importance of structures of non-integer stoichiometry or thermodynamical metastability in searching for novel compounds such as high temperature superconductors.

\begin{figure*}[!htbp]
    \centering
    \includegraphics[width=0.9\textwidth]{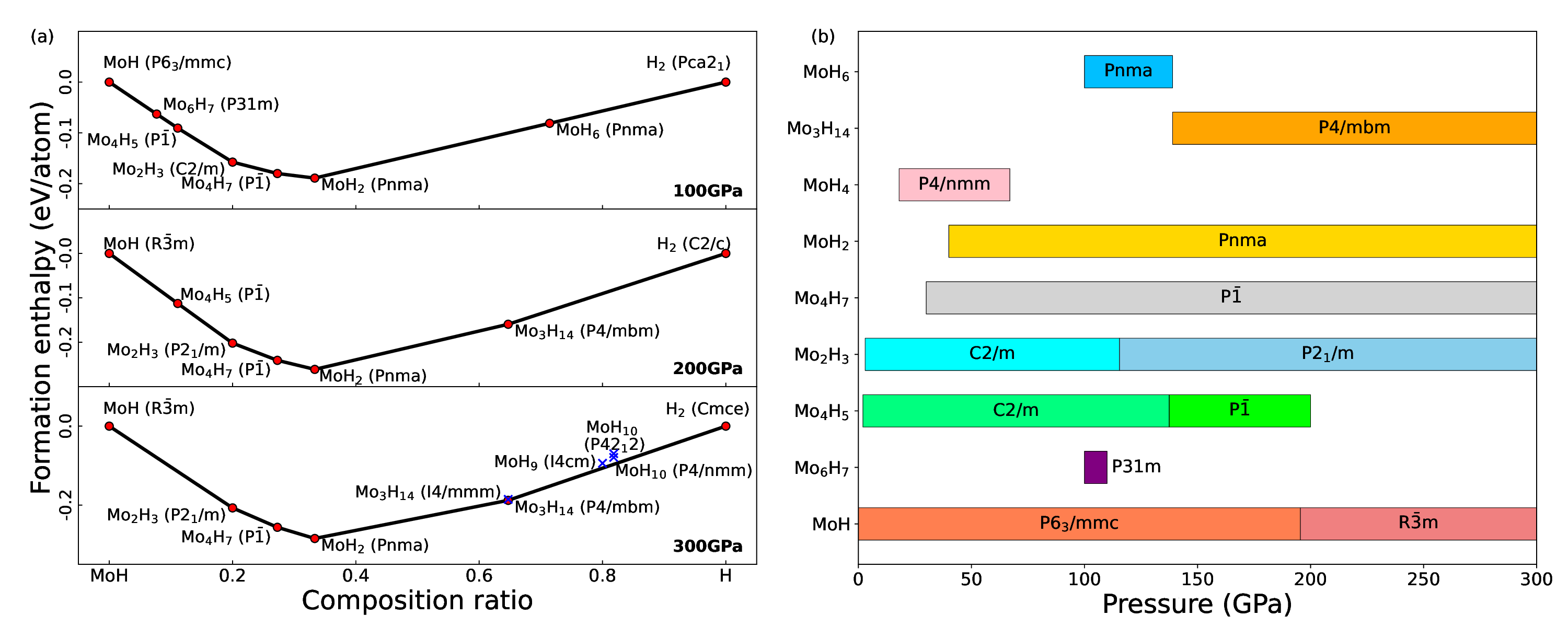}
    \caption{(a) Formation enthalpies of the Mo-H system at 100, 200 and 300 GPa. Red dots indicate thermodynamically stable phases and blue 'x' indicate metastable phases above the convex hull of black solid line. (b) The pressure-composition phase diagram of the Mo-H system in the pressure range of 0-300 GPa.}
    \label{CH}
\end{figure*}

\section{Method}

The evolutionary algorithm USPEX \cite{USPEX} is adopted in our structure prediction calculations. Totally more than 3000 structures are generated in our variable-composition searches in 50 generations, and more than 400 structures are generated in each fixed-compositions search in 8-12 generations. In both searches up to 40 atoms are included in the unit cell. The Vienna ab initio simulation program (VASP) \cite{VASP.PRB1996} was used in structure prediction calculations for structural relaxation and derivations of the total energy. The convergence criteria for force and energy are 3 meV/Å and 1$\times$10$^{-5}$ eV, repectively. We used projector augmented planewave (PAW) \cite{PAW} potentials with an energy cutoff of 500 eV and the $\Gamma$-centered k-point meshes of 2$\pi \times$ 0.06 \AA$^{-1}$ resolutions for Brillouin zone sampling to ensure the convergence of total energy. We also conducted higher precision calculations to pin down the convex hull and the composition-pressure phase diagram. More accurate convergence criteria for force and energy at 0.01 meV/Å and 1$\times$10$^{-9}$ eV were token. The PAW energy cutoff is increased to 800 eV and the k-point meshes of the Brillouin zone is 2$\pi \times$ 0.03 \AA$^{-1}$.

The electronic structures and the phonon related calculations are performed using the Quantum Espresso package \cite{Giannozzi_2009,Giannozzi_2017}. The optimized norm-conversing pseudopotentials \cite{Hamann_PhysRevB.88.085117,VANSETTEN201839} were used with the valence electrons 4s$^2$4p$^6$4d$^5$5s$^1$ for Mo and 1s$^1$ for H. The kinetic energy cutoff and the charge density cutoff for wavefunctions were set to 70 and 280 Ry, respectively. The Methfessel-Paxton \cite{MP_PhysRevB.40.3616} smearing with a spreading of 0.02 Ry was used for the Brillouin zone integration. A uniform $\Gamma$-centered k-point of 12$\times$12$\times$12 were employed in the electronic self-consistent calculations. The structures were optimized until the maximum energy and force were less than 1.0$\times$10$^{-7}$ Ry and 1.0$\times$10$^{-5}$ Ry/Bohr, respectively. Phonon dispersion and electron-phonon coupling (EPC) were calculated within density functional perturbation theory \cite{DFPT_RevModPhys.73.515}. Self-consistent electron density and electron-phonon coefficients were calculated with 24$\times$24$\times$24 k-point meshes and 4$\times$4$\times$4 q-point meshes. 

The superconductive transition temperatures were estimated using the McMillan equation, 
\begin{eqnarray}
T_c = \frac{\omega_{log}}{1.2} \exp \left ( -\frac{1.04(1+\lambda)}{\lambda - \mu^\ast\left (1+0.62 \lambda 
 \right ) }\right )
\end{eqnarray}
The electron-phonon coupling constant $\lambda$ and the logarithmic average frequency $\omega_{log}$ can be obtained from Eliashberg function $\alpha^2F(\omega)$ by
\begin{eqnarray}
\lambda  & = & 2 \int \frac{\alpha^2F(\omega)}{\omega} d\omega
\end{eqnarray}
and
\begin{eqnarray}
\omega_{log}  & = & \exp \left ( \frac{2}{\lambda} \int \frac{d\omega}{\omega} \alpha^2F(\omega)\ln(\omega) \right )
\end{eqnarray}

\section{Results and 
discussion}
\subsection{Convex hull and phase stabilities}

We have performed comprehensive variable-composition and fixed-compositions structure searches in the Mo-H system at the pressures of 100, 200 and 300 GPa. Comparing with previous studies, thermodynamically stable phases of non-integer H/Mo composition ratio are discovered in our study. All thermodynamically stable structures at the convex hull are marked by red dots, as shown in Fig.~\ref{CH}(a). At 100 GPa, there are seven thermodynamically stable phases, the new P31m-Mo$_6$H$_7$, C2/m-Mo$_4$H$_5$, C2/m-Mo$_2$H$_3$ and P$\bar{1}$-Mo$_4$H$_7$ structures and the previously known P6$_3$/mmc-MoH, Pnma-MoH$_2$ and Pnma-MoH$_6$ structures. \cite{Feng.SSC2016,Liao.JAP2020,Duan.PCCP2021} Based upon the output structures of the searches, we further derived the pressure-composition phase diagram of Mo-H system in the range of 0-300 GPa as summarized in Fig.~\ref{CH}(b). The Pnma-MoH$_2$ structure remains stable up to 300 GPa, consistent with previous work \cite{Liao.JAP2020,Duan.PCCP2021}. In contrast, MoH undergoes a structural phase transition from P6$_3$/mmc structure to R$\bar{3}$m structure at around 195 GPa. The Pnma-MoH$_6$ structure remains stable in the pressure range of 100-139 GPa, which is narrower than the range of 75-175 GPa predicted by Du et al. \cite{Duan.PCCP2021}.
Moreover, we found that P4/nmm-MoH$_4$ is stable in the range of 18-67 GPa. The I4/mmm-MoH$_4$ and the Immm-MoH$_6$ structures discovered by Liao et al. \cite{Liao.JAP2020} are thermodynamically metastable in our study. P31m-Mo$_6$H$_7$ structure is thermodynamically stable from 100 to 110 GPa. P4/mbm-Mo$_{3}$H$_{14}$ structure is thermodynamically stable above 140 GPa. The Cmmm-MoH$_{11}$ structure predicted by Du et al. \cite{Duan.PCCP2021} becomes thermodynamically metastable due to the appearance of P4/mbm-Mo$_3$H$_{14}$ phase in our work.
Mo$_2$H$_3$ of C2/m structure transforms into more energetically favorable P2$_1$/m structure at about 115 GPa and remains thermodynamically stable until 300 GPa. Mo$_4$H$_5$ of C2/m structure transforms into P$\bar{1}$ structure above 137 GPa, which remains stable until 200 GPa.

In addition to the thermodynamically stable structures, we also examined the superconductivity of energetic competitive metastable phases within 50 meV/atom above the convex hull. Totally nine metastable molybdenum-hydrides are found superconducting at 300 GPa, including P4$_2$/mnm-MoH$_3$ (5.7 meV/atom above hull), I4/m-MoH$_4$ (11.8 meV/atom above hull), I4/mmm-Mo$_3$H$_{14}$ (2.5 meV/atom above hull), P4/mmm-Mo$_3$H$_{14}$ (2.3 meV/atom above hull), R3-Mo$_3$H$_{14}$ (37.3 meV/atom above hull), P31m-Mo$_3$H$_{14}$ (39.6 meV/atom above hull), I4cm-MoH$_9$ (12.5 meV/atom above hull), P4/nmm-MoH$_{10}$ (20.6 meV/atom above hull) and P42$_1$2-MoH$_{10}$ (27.0 meV/atom above hull). As a matter of fact, metastable materials, from polymorphs of metals \cite{Turnbull1981,Badding1995,Bakker1995,Koinuma2004} to fullerene C$_{60}$ \cite{C60}, have long been synthesized and implemented. \cite{Aykol.sciadv2018} P6$_3$/mmc-NdH$_9$ is predicted 35 meV/atom above the convex hull at 150 GPa, and has been successfully synthesized at high pressure with T$_c$ $\approx$ 4.5 K. \cite{DZhou.jacs2020} Previous structural searches also predicted the metastable ternary Li$_2$MgH$_{16}$ of $\sim$20 meV/atom above the convex hull, which has a very high T$_c$ $\approx$ 473 K at 250 GPa \cite{SunY.PRL2019, PhysRevLett.128.186001}. Therefore, the metastable nature of these discovered molybdenum-hydrides in our study doesn't necessarily exclude the possibility for the synthesis of them in experiments. \cite{Wu.EES2013}

\subsection{Superconductivity, phonon spectra and electron-phonon coupling} 

\begin{table}[!ht]
    \caption{The calculated electronic DOS of H at the Fermi level N$_H$(E$_F$), the total DOS at the Fermi level N(E$_F$), the EPC parameter $\lambda$, the logarithmic average frequency $\omega_{log}$, and the superconducting transition temperature T$_c$ of molybdenum-hydrides at 300 GPa, where the thermodynamically stable structures are shown in bold.}
    \label{SCparams}
    \begin{tabular}{cccccccc}
    \hline
 Mo-H & Space group & N$_H$(E$_F$) & N(E$_F$) & $\lambda$ & $\omega_{log}$ & T$_c$ &\\
    && \multicolumn{2}{c}{(states/meV/\AA$^3$)} & & (cm$^{-1}$) & (K) &\\ 
\hline
    \textbf{MoH} & \textbf{R$\bar3$m} & \textbf{2.6} & \textbf{57.9} & \textbf{0.56} & \textbf{411} & \textbf{7-11} & \\ 
    MoH$_3$ & P4$_2$/mnm & 3.0 & 35.3 & 0.47 & 521 & 4-7 &  \\ 
    MoH$_4$ & I4/m & 5.8 & 47.2 & 0.65 & 545 & 17-23 & \\ 
    Mo$_3$H$_{14}$ & P4/mmm & 4.1 & 43.6 & 0.65 & 404 & 13-17 & \\ 
    \textbf{Mo$_3$H$_{14}$} & \textbf{P4/mbm} & \textbf{4.2} & \textbf{48.0} & \textbf{0.60} & \textbf{518} & \textbf{12-17} & \\ 
    Mo$_3$H$_{14}$ & P31m & 5.8 & 28.1 & 0.62 & 678 & 18-25 & \\ 
    Mo$_3$H$_{14}$ & R3 & 6.0 & 28.2 & 0.58 & 670 & 14-20 & \\ 
    Mo$_3$H$_{14}$ & I4/mmm & 16.4 & 60.9 & 1.09 & 597 & 60-68 &  \\ 
    MoH$_9$ & I4cm & 9.3 & 25.9 & 0.88 & 811 & 55-66 &\\ 
    MoH$_{10}$ & P4/nmm & 17.1 & 32.0 & 2.24 & 432 & 90-96 & \\ 
    MoH$_{10}$ & P42$_1$2 & 15.5 & 29.7 & 1.77 & 665 & 116-126 & \\ 
    \hline
    \end{tabular}
\end{table}

\begin{figure*}[!htbp]  
    \centering
    \includegraphics[width=0.9\textwidth]{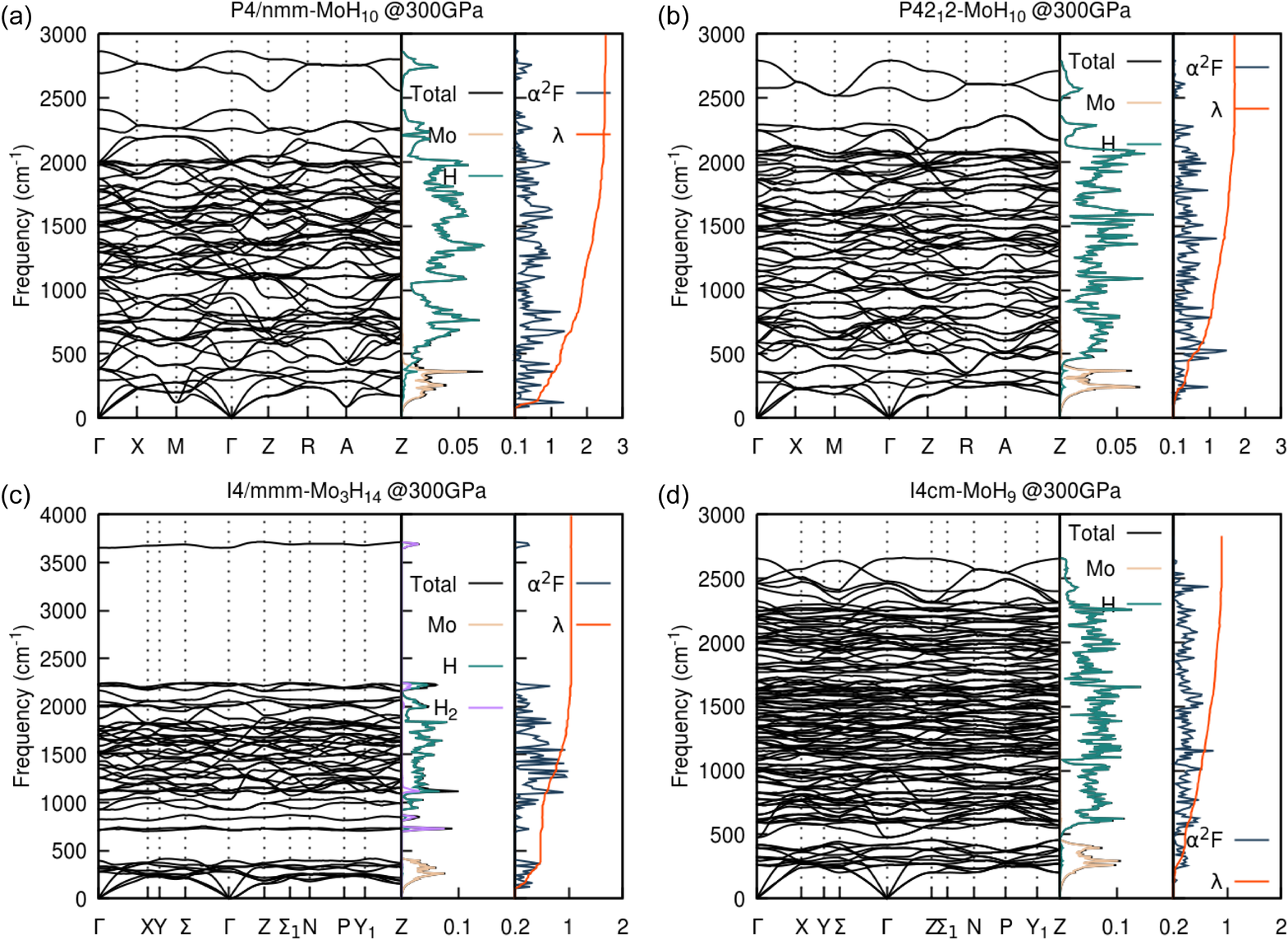}
    \caption{The phonon dispersion relation, the total and the projected PHDOS, the Eliashberg function $\alpha^2F(\omega)$ and the EPC parameter $\lambda$ of P4/nmm-MoH$_{10}$ (a) P42$_1$2-MoH$_{10}$ (b) I4/mmm-Mo$_3$H$_{14}$ (c) and I4cm-MoH$_9$ (d) at 300 GPa.}
    \label{ph}
\end{figure*}

The superconducting transition temperature T$_c$ at 300 GPa of all eleven new structures discovered in this study are listed in Table~\ref{SCparams}, together with their total electronic density of states (DOS) at the Fermi level N(E$_F$), the hydrogen-DOS at the Fermi level N$_H$(E$_F$), the EPC parameter $\lambda$ and the logarithmic average frequency $\omega_{log}$. Among all thermodynamically stable structures, only R$\bar3$m-MoH and P4/mbm-Mo$_3$H$_{14}$ are found superconducting with transition temperature at 7-11 K and 12-17 K, respectively. The hydrogen-DOS at the Fermi level N$_H$(E$_F$), the EPC parameter $\lambda$, and the superconducting temperatures T$_c$ of I4/mmm-Mo$_3$H$_{14}$, I4cm-MoH$_9$, P4/nmm-MoH$_{10}$ and  P42$_1$2-MoH$_{10}$ are significantly larger than these of other seven structures. Obviously, in our study the hydrogen-DOS at the Fermi level N$_H$(E$_F$) play critical roles in control the EPC parameter $\lambda$ and consequently the superconducting transition temperature T$_c$ of molybdenum hydrides. In order to explain the origin of high-temperature superconductivity in molybdenum hydrides, we mainly focus on these four superconducting structures, I4/mmm-Mo$_3$H$_{14}$, I4cm-MoH$_9$, P4/nmm-MoH$_{10}$ and P42$_1$2-MoH$_{10}$, in later contents.

The phonon spectra and the phonon density of states (PHDOS) of P4/nmm-MoH$_{10}$, P42$_1$2-MoH$_{10}$, I4/mmm-Mo$_3$H$_{14}$ and I4cm-MoH$_9$ at 300 GPa are shown in Fig.~\ref{ph}. Absence of imaginary frequency modes in the whole Brillouin zone suggests the dynamical stability of these phases. As shown in the projected PHDOS, the phonon modes of these phases can be classified into two parts, the high frequency modes originating from vibrations of H-atoms above 500 cm$^{-1}$ and the low frequency modes being dominated by vibrations of Mo-atoms below 500 cm$^{-1}$. 

\begin{figure*}[!htbp]
    \centering
    \includegraphics[width=0.8\textwidth]{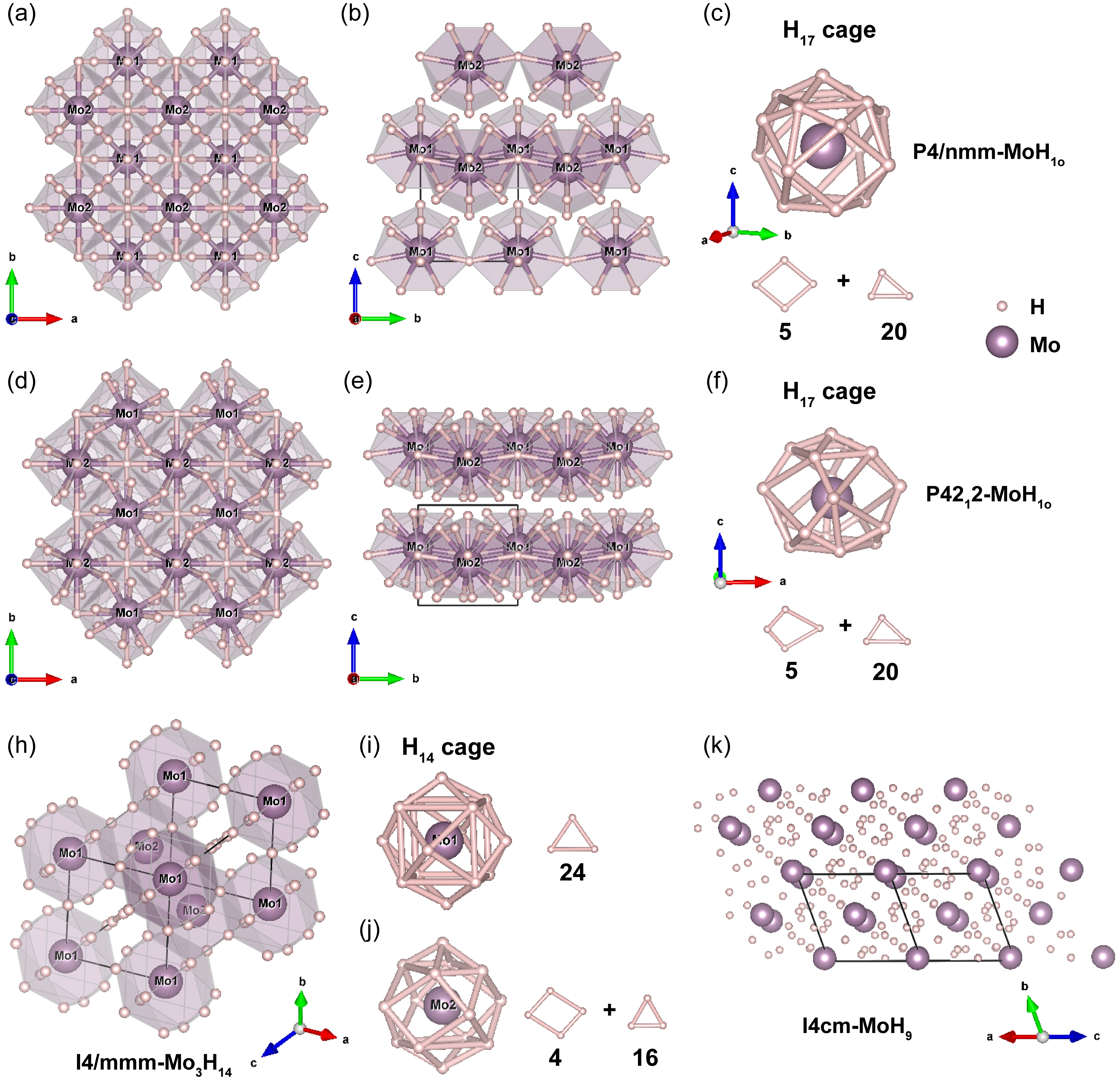}
    \caption{Crystal structures of P4/nmm-MoH$_{10}$ (a, b) P42$_1$2-MoH$_{10}$ (d, e) I4/mmm-Mo$_3$H$_{14}$ (h) and I4cm-MoH$_9$ (k) at 300 GPa. (c) and (f) depict the Mo-centered H$_{17}$ cage of P4/nmm-MoH$_{10}$ and P42$_1$2-MoH$_{10}$, respectively. Each H$_{17}$ cage contains 5 quadrilateral and 20 triangles. (i) and (j) depicts the Mo-centered H$_{14}$ cage in I4/mmm-Mo$_3$H$_{14}$. The Mo1-H$_{14}$ cage contains 24 isosceles triangle faces, and the Mo2-H$_{14}$ cage contains 4 irregular quadrilaterals and 16 isosceles triangles. The small and large spheres represent H- and Mo-atoms, respectively.}
    \label{cs}
\end{figure*}

The Eliashberg function $\alpha^2F(\omega)$ and the accumulated EPC parameter $\lambda$ are presented in the rightmost panel of each subfigure of Fig.~\ref{ph}. The vibrations of Mo-atoms at the low-frequency region contribute about 53\%, 23\%, 45\% and 24\% of the total $\lambda$ for P4/nmm-MoH$_{10}$, P42$_1$2-MoH$_{10}$, I4/mmm-Mo$_3$H$_{14}$ and I4cm-MoH$_9$, respectively. This indicates that the low-frequency vibrations of Mo-atoms plays a non-negligible role in enhancing the superconducting temperature, especially for P4/nmm-MoH$_{10}$ and I4/mmm-Mo$_3$H$_{14}$ structures. The resulting EPC parameters $\lambda$ of P4/nmm- and P42$_1$2-MoH$_{10}$ are quite large at values of 2.24 and 1.77 at 300 GPa. With typical Coulomb pseudopotential parameters $\mu^*$ from 0.1 to 0.13, T$_c$ of P4/nmm- and P42$_1$2-MoH$_{10}$ reach 90-96 K and 116-126 K at 300 GPa, respectively. For P42$_1$2-MoH$_{10}$, despite the relatively small $\lambda$ compared to P4/nmm-MoH$_{10}$, it has even higher T$_c$ owing to its larger $\omega_{log}$ at 665 cm$^{-1}$. 

For I4/mmm-Mo$_3$H$_{14}$, a noticeable feature of the phonon spectrum is the flat dispersion related to the vibrations of the H$_2$ molecule at very high frequencies (above 3500 cm$^{-1}$). Although the H-DOS of I4/mmm-Mo$_3$H$_{14}$ is comparable to that of P42$_1$2-MoH$_{10}$ as presented in Table~\ref{SCparams}, its T$_c$ is estimated to be only 60-68 K due to the relatively weak electron-phonon coupling.

The distribution of phonon frequencies in I4cm-MoH$_9$ is relatively narrower. However, it has a logarithmic average frequency $\omega_{log}$ at 811 cm$^{-1}$, which indicate that high-frequency phonons contribute significantly to the EPC strength. Although the H-DOS at the Fermi level N$_H$(E$_F$) and the EPC strength of I4cm-MoH$_9$ are the smallest among the four high-temperature superconducting phases, at 9.3 state/meV/\AA$^3$ and 0.88, the estimated upper limit of the superconducting temperature T$_c$ of I4cm-MoH$_9$ is 66 K owing to the large $\omega_{log}$.

\begin{figure*}[!htbp]
    \centering
    \includegraphics[width=0.9\textwidth]{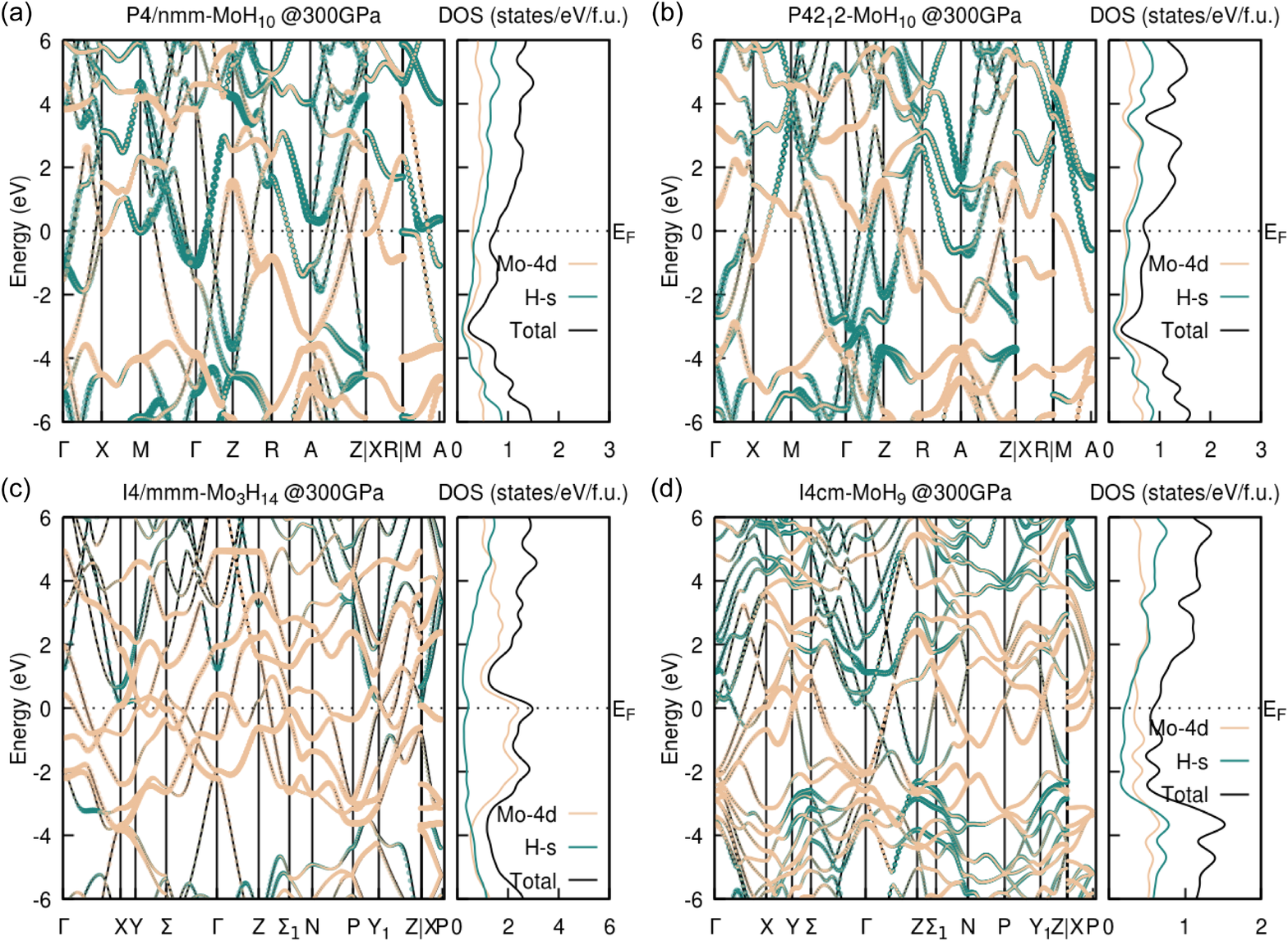}
    \caption{The electronic band structure and density of states (DOS) of P4/nmm-MoH$_{10}$ (a) P42$_1$2-MoH$_{10}$ (b) I4/mmm-Mo$_3$H$_{14}$ (c) and I4cm-MoH$_9$ (d) at 300 GPa.}
    \label{elec}
\end{figure*}

\subsection{Crystal structures and electronic properties}

The crystal structures of P4/nmm-MoH$_{10}$, P42$_1$2-MoH$_{10}$, I4/mmm-Mo$_3$H$_{14}$ and I4cm-MoH$_9$ at 300 GPa are shown in Fig.~\ref{cs}. The detailed structure parameters are listed in Table.S1 of the Supplementary Information. For P4/nmm-MoH$_{10}$ and P42$_1$2-MoH$_{10}$, the two crystal structures are very close. In both structure, each Mo-atom has 17 nearest neighbour H-atoms that form a cage surrounding the Mo-atom at the center. The H$_{17}$ cage consists of 25 faces, including 5 irregular quadrilaterals and 20 triangles (see Fig.~\ref{cs}(c) and (f)). Two adjacent hydrogen cages share a quadrilateral face, and four neighbour cages share a vertex. In the Mo-H$_{17}$ cages of P4/nmm-MoH$_{10}$ and P42$_1$2-MoH$_{10}$, the nearest H-H distances are 1.15 and 1.08 Å, and the average Mo-H distances are 1.72 and 1.73 Å, respectively.

In I4/mmm-Mo$_3$H$_{14}$, each Mo-atom is surrounded by a hydrogen cage consists of 14 nearest neighbour H-atoms. There are two types of H$_{14}$ cages in I4/mmm-Mo$_3$H$_{14}$, the one enclosing the Mo1-atom has 24 isosceles triangles (see Fig.~\ref{cs}(i)), and the other enclosing the Mo2-atom has 4 irregular quadrilaterals and 16 isosceles triangles (see Fig.~\ref{cs}(j)). The adjacent Mo2-H$_{14}$ cages share an irregular quadrilateral face. In particular, I4/mmm-Mo$_3$H$_{14}$ structure contains two H$_2$ molecular like units with a bond length of 0.765 \AA at 300 GPa. Different from the calthrate structures in P4/nmm-MoH$_{10}$, P42$_1$2-MoH$_{10}$ and I4/mmm-Mo$_3$H$_{14}$, I4cm-MoH$_9$ crystal is composed of alternating Mo- and H-layers, as shown in Fig.~\ref{cs}(k).

The electronic band structures and density of states of P4/nmm-MoH$_{10}$, P42$_1$2-MoH$_{10}$, I4/mmm-Mo$_3$H$_{14}$ and I4cm-MoH$_9$ at 300 GPa are shown in Fig.~\ref{elec}. In all four high-T$_c$ molybdenum-hydrides, both Mo-4d and H-s have significant contributions to the total DOS near the Fermi level. In particular, the H-s DOS in P4/nmm-MoH$_{10}$ and P42$_1$2-MoH$_{10}$ are comparable to the Mo-4d DOS. Previous experience in the rare-earth hydrides is that the transfer of rare-earth-4f electrons to the H-s orbitals would enlarge the effective number of H-s electrons participating the EPC and then enhance the superconducting transition temperature T$_c$.
In electronic band structures of P4/nmm-MoH$_{10}$ and P42$_1$2-MoH$_{10}$, the Mo-4d electrons and the H-s electrons are strongly hybridized around the Fermi level (see Fig.~\ref{elec}(a, b)), which suggests the transfer of Mo-4d electrons to the H-s orbitals. It explains the large H-s DOS and the high superconducting transition temperature of P4/nmm-MoH$_{10}$ and P42$_1$2-MoH$_{10}$. In contrast, in I4/mmm-Mo$_3$H$_{14}$ and I4cm-MoH$_9$ the hybridization between Mo-4d and H-s orbitals around the Fermi level is less pronounced as shown in Fig.~\ref{elec}(c) and (d), which is consistent with their relatively smaller H-s DOS at the Fermi level and their relatively lower superconducting transition temperature. There is no van Hove singularity in the electronic DOS of these superconducting hydrides in our study.

\subsection{Role of H\texorpdfstring{$_2$} dumbbells}

There are H$_2$-like units in I4/mmm-Mo$_3$H$_{14}$, P4/mmm-Mo$_3$H$_{14}$ and P4/mbm-Mo$_3$H$_{14}$ phases. The presence of H$_2$-like units were also observed in previous structrual searches in MoH$_{11}$, NbH$_{11}$, TaH$_{11}$ and WH$_{11}$ of Cmmm structure. \cite{Duan.PCCP2021} Established experience about superconducting hydrides with hydrogen clathrate structures tells us that H$_2$-like unit doesn't help in enhancing the superconducting transition temperature, since it will not contribute to either the H-DOS at the Fermi level N$_H$(E$_F$) or the EPC parameterized by $\lambda$. \cite{Peng.PRL2017} In our study of I4/mmm-Mo$_3$H$_{14}$, P4/mmm-Mo$_3$H$_{14}$ and P4/mbm-Mo$_3$H$_{14}$, N$_H$(E$_F$) is 16.4, 4.1, 4.2 states/meV/\AA$^3$ and $\lambda$ is 1.09, 0.65, 0.60, as shown in Table.~\ref{SCparams}. As a result, the superconducting transition temperature T$_c$ of I4/mmm-Mo$_3$H$_{14}$ is 60-68 K, much higher than the T$_c$ of P4/mmm-Mo$_3$H$_{14}$ and P4/mbm-Mo$_3$H$_{14}$ at 13-17 K and 12-17 K, respectively. 

On the other hand, weakening or even elimination of the H-H bonding will help promoting the T$_c$ of hydrides with hydrogen clathrate structures. The H-H bond length in I4/mmm-Mo$_3$H$_{14}$ equals to 0.765 \AA, which is longer than the H-H bond lengths of P4/mmm-Mo$_3$H$_{14}$ and P4/mbm-Mo$_3$H$_{14}$ at 0.736 and 0.744 Å, respectively. The relatively longer H-H bond length of I4/mmm-Mo$_3$H$_{14}$ suggests suppressed bonding of the H$_2$-like unit, as well as enhanced N$_H$(E$_F$), $\lambda$ and T$_c$. By introduction of extra charges, for example replacing the Mo-atom by heavier W-atom, the strength of H-H bond in I4/mmm-Mo$_3$H$_{14}$, P4/mmm-Mo$_3$H$_{14}$ and P4/mbm-Mo$_3$H$_{14}$ could be suppressed, therefore the H-DOS at the Fermi level, the EPC parameter, and the superconducting transition temperature could be enhanced. 

\section{Conclusion}
In summary, we have performed comprehensive variable-composition structural searches for Mo-H binary system at pressures up to 300 GPa. We reported an updated phase diagram of the thermodynamically stable structures in Mo-H, including structures of five unprecedented non-integer H/Mo composition ratios, Mo$_2$H$_3$, Mo$_4$H$_5$, Mo$_6$H$_7$, Mo$_4$H$_7$ and Mo$_{3}$H$_{14}$. Besides all previously known superconducting molybdenum hydrides, series of thermodynamically metastable superconducting structures are identified, in which I4/mmm-Mo$_3$H$_{14}$, I4cm-MoH$_9$, P4/nmm-MoH$_{10}$ and P42$_1$2-MoH$_{10}$ have relatively high superconducting transition temperatures from 55 to 126 K at 300 GPa. Further analysis indicates that high-T$_c$ of these superconducting molybdenum hydrides originates from collaborating contributions by both the H-atoms and the Mo-atoms. Our results confirm that in superconducting hydrides with hydrogen clathrate structures, H$_2$-like unit plays negative role for elevating the T$_c$. This research underlines the importance of phases of non-integer composition ratio or metastable states in structural searches, therefore sheds new lights on future explorations of materials with novel properties, such as high temperature superconductors.

\section*{Acknowledgment}
This work was supported by the National Natural Science Foundation of China (11604255) and the Natural Science Basic Research Program of Shaanxi (2021JM-001). The calculations are performed at the HPC platform of Xi'an Jiaotong University.

\bibliography{bib}

\end{document}